\documentclass[letterpaper,notoc]{CECS}

\title{{\bf Finite action principle for Chern-Simons AdS gravity}}
\author{P. Mora$^{1}$, R. Olea$^{2}$, R. Troncoso$^{3}$ and J. Zanelli$^{3}$ \\
$^{1}$Instituto de F\'{\i}sica, Facultad de Ciencias Igu\'{a} 4225, Montevideo, Uruguay.\\
$^{2}$Departamento de F\'{\i}sica, Pontificia Universidad
Cat\'{o}lica de
Chile, Casilla 306, Santiago 22, Chile.\\
$^{3}$Centro de Estudios Cient\'{\i}ficos (CECS), Casilla 1469,
Valdivia, Chile.}

\preprint{{\tiny CECS-PHY-04/07} }

\abstract{ A finite action principle for Chern-Simons AdS gravity
is presented. The construction is carried out in detail first in
five dimensions, where the bulk action is given by a particular
combination of the Einstein-Hilbert action with negative
cosmological constant and a Gauss-Bonnet term; and is then
generalized for arbitrary odd dimensions. The boundary term needed
to render the action finite is singled out demanding the action to
attain an extremum for an appropriate set of boundary conditions.
The boundary term is a local function of the fields at the
boundary and is sufficient to render the action finite for
asymptotically AdS solutions, without requiring background fields.
It is shown that the Euclidean continuation of the action
correctly describes black hole thermodynamics in the canonical
ensemble. Additionally, background independent conserved charges
associated with the asymptotic symmetries can be written as
surface integrals by direct application of Noether's theorem.}
\begin{document}
\section{Introduction}

The relevance of having a finite action for gravity has been recognized since
the early days of black hole thermodynamics \cite{Gibbons-Hawking}. In the
presence of a negative cosmological constant, this problem has gained new
interest in view of the AdS/CFT correspondence (see \cite{MAGOO} and references
therein). A first approach to regularize the action involved choosing a
reference background with suitable matching conditions \cite {Hawking-Page}.
However, as these conditions depend on the topology and the asymptotic behavior
of the solution, the background must be selected case by case. Moreover, such a
choice is not necessarily unique and in some cases, it might prove impossible
to find one.

An interesting alternative to tackle this problem, inspired by the AdS/CFT
correspondence, is the background independent counterterms method \cite
{Henningson-Skenderis} (see also \cite{Balasubramanian-Kraus}), which
involves the addition of a linear combination of invariants of the intrinsic
geometry at the boundary. However, as the spacetime dimension increases
there is a plethora of possible terms (see e.g., \cite{Kerr-AdS,NUT-AdS}),
and the full series for any dimension is unknown.

On the other hand, in dimensions greater than four, gravity can be described by
a generalization of the Einstein-Hilbert action, whose Lagrangian is an
arbitrary linear combination of the dimensional continuation of the Euler
densities \cite{LL,Zumino}. These so-called Lovelock Lagrangians still lead to
second order field equations for the metric. In odd dimensions there exist
special combinations of these terms that allows to write the Lagrangians as
Chern-Simons densities \cite{Chamseddine-G} which possess a number of
interesting features. In particular, apart from general coordinate invariance,
these theories possess an enhanced local symmetry and their local
supersymmetric extensions have been constructed in three \cite
{Achucarro-Townsend,HIPT}, five \cite{Chamseddine-Sugra}, and higher odd
dimensions \cite{BTZ,TZ11,TZ-B,HDG,HOT,HTZ}.

Here we discuss an action principle for Chern-Simons AdS gravity which is
background independent for all odd dimensions. As both the background and the
counterterm approaches described above become cumbersome when one deals with a
Lagrangian containing higher powers in the curvature in higher odd dimensions,
we follow a different strategy.

A set of boundary conditions for the geometry is proposed, which makes the
Chern-Simons action functionally differentiable at the extremum. The boundary
conditions single out a boundary term which is a local function of the fields
at the boundary. The resulting action principle requires no background fields,
is finite for configurations that satisfy the boundary conditions\textbf{,} and
the Euclidean continuation of the action correctly describes the thermodynamics
of the system in the canonical ensemble. Furthermore, conserved charges
associated with the asymptotic symmetries can be written as surface integrals
at infinity by direct application of Noether's theorem without the need for ad
hoc regularizations.

In the next section the five dimensional case is analyzed in detail. Section 3
is devoted to generalize the results for any odd dimension. The applications to
black hole thermodynamics, and the construction of the conserved charges is
carried out in Sections 4 and 5, respectively.

\section{Chern-Simons-AdS Gravity in Five Dimensions}

Let us consider five-dimensional gravity with a negative cosmological
constant and a Gauss-Bonnet term
\begin{eqnarray}
I_{5} &=&-\frac{4\kappa }{l^{2}}\int\limits_{M}d^{5}x\sqrt{-g}\left[ R+\frac{%
6}{l^{2}}+\frac{l^{2}}{4}\left( R_{\mu \nu \alpha \beta }R^{\mu \nu \alpha
\beta }-4R_{\mu \nu }R^{\mu \nu }+R^{2}\right) \right]+\int\limits_{\partial M}B_{4}\;,  \label{GB}
\end{eqnarray}
where the relative coefficients in (\ref{GB}) have been fixed so that there is
a unique negative constant curvature solution (AdS) with radius $l$. The
boundary term $B_{4}$ can be found as follows. We require the action to have an
extremum under variations of the geometry with a fixed extrinsic curvature at
the boundary, i.e.,
\begin{equation}
\delta K_{ij}=0\;,  \label{deltaK}
\end{equation}
where $K_{ij}$ is fixed in terms of the boundary metric $h_{ij}$ as
\begin{equation}
K_{ij}=\Omega h_{ij}\;,  \label{K=g}
\end{equation}
for some arbitrary function $\Omega $ at $\partial M$. The remarkable feature
of these boundary conditions is that the boundary term can be integrated to be
\begin{eqnarray}
B_{4} &=&2\kappa \sqrt{-h}\left[ -K\left( \frac{1}{l^{2}}+\widetilde{R}+\frac{1}{%
2}\left( 3K_{ij}K^{ij}-K^{2}\right) \right) \right.  \nonumber \\
&&\qquad \qquad +\left. 4\widetilde{R}_{ij}K^{ij}+K_{j}^{i}K_{k}^{j}K_{i}^{k}%
\right] \;,  \label{B4Tensorial}
\end{eqnarray}
so that the resulting action principle is background independent.
Furthermore, this procedure also ensures the convergence of the action for
asymptotically AdS\ configurations. In Eq. (\ref{B4Tensorial}), $\widetilde{R}%
_{ij}$, and $\widetilde{R}$ are the Ricci tensor and the curvature scalar of
$\partial M$, respectively. The extrinsic curvature is the Lie derivative along
the normal to the boundary, $n$,
\begin{equation}
K_{ij}={\cal L}_{n}h_{ij}\;.  \label{KijLie}
\end{equation}
Hence, the condition (\ref{K=g}) means that a diffeomorphism along $n$ is
related to a conformal transformation for the metric at $\partial M$. Thus,
the boundary condition (\ref{K=g}) is in this sense ``holographic''\footnote{%
A submanifold that satisfies this condition is also known as {\em totally
umbilical} \cite{Spivak}}. One of the differences between our approach and the
counterterms method of \cite{Henningson-Skenderis,Balasubramanian-Kraus} is
that here, instead of fixing the metric at the boundary, we fix the extrinsic
curvature. This allows us to readily generalize our results for all odd
dimensions.

 In order to show this construction explicitly, as well
as to deal with this problem in higher dimensions, it is useful to employ the
first order formalism in terms of the spin connection $\omega ^{ab}=\omega
_{\mu }^{ab}dx^{\mu }$, and the vielbein $e^{a}=e_{\mu }^{a}dx^{\mu }$. In this
language, the action (\ref{GB}) reads
\begin{equation}
I_{5}=\kappa \int\limits_{M}\epsilon _{abcde}R^{ab}R^{cd}e^{e}+\frac{2}{%
3l^{2}}R^{ab}e^{c}e^{d}e^{e}+\frac{1}{5l^{4}}e^{a}e^{b}e^{c}e^{d}e^{e}+dB_{4}%
,  \label{I5}
\end{equation}
where the wedge products between forms is understood, and $R^{ab}\equiv
d\omega ^{ab}+\omega^a_{\,c}\omega ^{cb}$ is the curvature
two-form. Henceforth the AdS radius is chosen to be $l=1$. The Lagrangian
can be seen as the Chern-Simons form for the AdS group in five dimensions
\cite{Chamseddine-G}.

The variation of (\ref{I5}) yields the field equations plus a surface term $%
\Theta $

\begin{equation}
\delta I_{5}=\kappa \int\limits_{M}\mathcal{E}_{a}\delta e^{a}+2\mathcal{E}
_{ab}\delta \omega ^{ab}+\int\limits_{\partial M}\Theta  \label{varCSAdS5}
\end{equation}
with
\begin{eqnarray}
\mathcal{E}_{a} &=&\epsilon _{abcde}\bar{R}^{ab}\bar{R}^{de}\;,  \label{ea}
\\
\mathcal{E}_{ab} &=&\epsilon _{abcde}\bar{R}^{cd}T^{e}\;,  \label{eab}
\end{eqnarray}
where $\bar{R}^{ab}:= R^{ab}+e^{a}e^{b}$, and $T^{a}=de^{a}+\omega
_{b}^{a}e^{b}$ is the torsion. Clearly, $\Theta $ contains two parts

\begin{equation}
\Theta =\alpha _{4}+\delta B_{4}  \label{theta}
\end{equation}
where $\alpha _{4}$ is the boundary term coming from the variation of the
bulk action,
\begin{equation}
\alpha _{4}=2\kappa \epsilon _{abcde}\delta \omega ^{ab}e^{c}\left( R^{de}+%
\frac{1}{3}e^{d}e^{e}\right) .  \label{BTvar5}
\end{equation}

Thus, Eq. (\ref{theta}) can be expressed as

\begin{equation}
\Theta =2\kappa \int\limits_{\partial M}\epsilon _{abcdf}\delta \theta
^{ab}e^{c}\left( \widetilde{R}^{df}+\left( \theta ^{2}\right) ^{df}+\frac{1}{3}%
e^{d}e^{f}\right) +\delta B_{4}, \label{Theta}
\end{equation}
where $\left( \theta ^{2}\right) ^{ab}=\theta _{\;c}^{a}\theta ^{cb}$ (see
Appendix A.1). Here $ \theta ^{ab}$ is the second fundamental form which plays
the role of the extrinsic curvature, and $\widetilde{R}^{ab}$ is the intrinsic
curvature of the
boundary, which doesn't have normal components ({\em i.e.} $\widetilde{R}%
^{ab}n_{b}=0$), and is related to the curvature $R^{ab}$ through the
Gauss-Codazzi equations (see Appendix A.1).

\subsection{Boundary Conditions}

The action attains an extremum only if $\Theta $ vanishes, which determines the
variation of the boundary term $\delta B_{4}$. We look for a boundary condition
that allows to obtain $B_{4}$ as a local function of the fields at the boundary
without using background fields. A boundary condition that satisfies these
requirements is
\begin{equation}
\delta \theta ^{[ab}e^{c]}=\theta ^{[ab}\delta e^{c]}\;,  \label{BC}
\end{equation}
and it can be checked that the boundary conditions (\ref{deltaK}) with (\ref
{K=g}) are included in this set. In other words, the boundary conditions (%
\ref{deltaK}) with (\ref{K=g}), are a sufficient condition for the boundary
condition (\ref{BC}) to be satisfied. By appropriately splitting the integrand
in (\ref{Theta}), and using Eq. (\ref{BC}), $\delta B_{4}$ can be written as
\begin{eqnarray}
\delta B_{4} &=&-\int\limits_{\partial M}\kappa \epsilon _{abcde}\left[ \delta
\theta ^{ab}e^{c}\left( \widetilde{R}^{de}+\frac{3}{2}\left( \theta
^{2}\right) ^{de}+\frac{1}{6}e^{d}e^{e}\right) \right.  \nonumber \\
&&+\left. \theta ^{ab}\delta e^{c}\left( \widetilde{R}^{de}+\frac{1}{2}\left(
\theta ^{2}\right) ^{de}+\frac{1}{2}e^{d}e^{e}\right) \right] , \label{split}
\end{eqnarray}
which can be readily integrated as \cite{reshuffling}

\begin{equation}
B_{4}=-\kappa \epsilon _{abcde}\theta ^{ab}e^{c}\left( \widetilde{R}^{de}+\frac{1%
}{2}\left( \theta ^{2}\right) ^{de}+\frac{1}{6}e^{d}e^{e}\right) ,
\label{BT4}
\end{equation}
and in components reduces to (\ref{B4Tensorial}).

To summarize, the action principle in five dimensions is given by (\ref{I5}),
with the boundary term (\ref{BT4}). The variation of the action evaluated on a
solution of the field equations is
\begin{equation}
\delta I_{5}=\kappa \int\limits_{\partial M}\epsilon _{abcde}\left( \delta
\theta ^{ab}e^{c}-\theta ^{ab}\delta e^{c}\right) \left( \tilde{R}^{de}+%
\frac{1}{2}\left( \theta ^{2}\right) ^{de}+\frac{1}{2}e^{d}e^{e}\right)
,  \label{deltaIg5}
\end{equation}
which explicitly shows that it attains an extremum for the boundary
conditions (\ref{BC}).

As it is shown in the next section, this boundary condition ensures the
existence of a boundary term which is a local function of the fields at the
boundary, without using background fields, so that the higher-dimensional
Chern-Simons AdS action does have an extremum on-shell. The well-defined action
principle obtained so leads to the correct results for the black hole
thermodynamics, and also allows to express the conserved charges as surface
integrals in a straightforward way.

\section{Generalization to $d=2n+1$ dimensions}

A Chern-Simons action for gravity in $2n+1$ dimensions that generalizes (\ref
{GB}) is a linear combination of the form

\begin{equation}
I_{2n+1}=\kappa \int\limits_{M}\sum\limits_{p=0}^{n}\alpha
_{p}L^{(p)}+\int\limits_{\partial M}B_{2n},  \label{Lovelock}
\end{equation}
where $\kappa =(2\left( d-2\right) !\Omega _{d-2}G_{n})^{-1}$, and $L^{(p)}$
are the dimensional continuations of Euler densities from lower dimensions,

\[
L^{(p)}=\epsilon
_{a_{1}...a_{d}}R^{a_{1}a_{2}}...R^{a_{2p-1}a_{2p}}e^{a_{2p+1}}...e^{a_{d}}%
,
\]
and
\begin{equation}
\alpha _{p}=\frac{l^{2(p-n)}}{d-2p}\left(
\begin{array}{l}
n \\
p
\end{array}
\right)  \label{alphap}
\end{equation}

It is useful to rewrite the series (\ref{Lovelock}) in terms of an integral
over the continuous parameter $t$
\begin{equation}
I_{2n+1}=\kappa \int\limits_{M}\int\limits_{0}^{1}dt\;\epsilon
_{a_{1}...a_{2n+1}}R_{t}^{a_{1}a_{2}}...R_{t}^{a_{2n-1}a_{2n}}e^{a_{2n+1}}+%
\int\limits_{\partial M}B_{2n}  \label{LCSAdS}
\end{equation}
where $R_{t}^{ab}:= R^{ab}+t^{2}e^{a}e^{b}$. Following the same procedure as in
five dimensions the variation of the boundary term $B_{2n}$ in all odd
dimensions is found to be
\begin{equation}
\delta B_{2n}=-\kappa n\int\limits_{0}^{1}dt\;\epsilon \delta \theta e\left(
\widetilde{R}+\theta ^{2}+t^{2}e^{2}\right) ^{n-1},
\label{alphanoind}
\end{equation}
where we have simplified the notation omitting the indices, which are hereafter
assumed to be contracted in the canonical way. Using the same set of boundary
conditions defined in Eq. (\ref{BC}), the boundary term for Chern-Simons AdS
gravity in all odd dimensions is found to be
\begin{equation}
B_{2n}=-\kappa n\int\limits_{0}^{1}dt\int\limits_{0}^{t}ds\;\epsilon \theta
e\left( \widetilde{R}+t^{2}\theta ^{2}+s^{2}e^{2}\right) ^{n-1}.  \label{B2n}
\end{equation}
With this formula for the boundary term, the finite action (\ref{LCSAdS}) now
reads,
\begin{equation}
I_{2n+1}=\kappa \int\limits_{M}\int\limits_{0}^{1}dt\;\epsilon R_{t}^{n}e^{a_{2n+1}}+%
\int\limits_{\partial M}B_{2n} \;.  \label{FiniteAdSAction}
\end{equation}
It is straightforward to check that the variation of $I_{2n+1}$, evaluated on a
solution of the field equations,
\[
\delta I_{2n+1}=\kappa n\int\limits_{\partial
M}\int\limits_{0}^{1}dt\,t\;\epsilon \left( \delta \theta e-\theta \delta
e\right) \left( \widetilde{R}+t^{2}\theta ^{2}+t^{2}e^{2}\right) ^{n-1},
\]
vanishes for the boundary conditions (\ref{BC}).

It turns out that the action principle proposed here is finite for
configurations that satisfy the boundary conditions and the Euclidean
continuation of the action correctly describes the thermodynamics of the
system in the canonical ensemble, as it can be seen in the next section.

\section{Applications to Black Hole Thermodynamics}

Chern-Simons-AdS gravity possesses static black hole solutions, that are
asymptotically locally AdS, and whose line element is \cite{DCBH,CaiDC,aros}

\begin{equation}
ds^{2}=-\Delta ^{2}(r)dt^{2}+\frac{dr^{2}}{\Delta ^{2}(r)}+r^{2}d\Sigma
_{d-2}^{2}  \label{BH}
\end{equation}
with
\begin{equation}
\Delta ^{2}(r)=\gamma -\sigma +r^{2}.  \label{deltatop}
\end{equation}
Here $d\Sigma _{d-2}^{2}$ is the line element of the $(d-2)$-dimensional base
manifold\footnote{Requiring the existence of asymptotic Killing spinors
restricts the base manifold to be an Einstein space of Euclidean signature
satisfying $R_{mn}=\gamma (d-3)g_{mn}$ admitting at least one Killing spinor
\cite{AMTZ}. These manifolds are classified, and the constant $\gamma $ can be
normalized to $\pm 1,0$ by a suitable coordinate rescaling.} whose curvature
is related to $\gamma =1,0,-1$, and the horizon is located at $r_{+}=\sqrt{%
\sigma -\gamma }$. Requiring smoothness of the Euclidean black hole solution
at the horizon fixes the period of the Euclidean time as

\begin{equation}
\beta =T^{-1}=\left( \frac{1}{4\pi }\left. \frac{d\Delta ^{2}}{dr}\right|
_{r_{+}}\right) ^{-1}=\frac{2\pi }{r_{+}}  \label{beta}
\end{equation}
where $T$ is the black hole temperature.

In the semiclassical approximation, the partition function is given by $%
Z\approx e^{I_{2n+1}^{E}}$, where $I_{2n+1}^{E}$ is the Wick-rotated version of
the action (\ref{FiniteAdSAction}). For fixed temperature, the Euclidean action
is related to the free energy in the canonical ensemble, $I_{E}=-\beta
F=S-\beta M$, which defines the mass and entropy of the black hole. For the
spherically symmetric case ($\gamma =1$), the boundary term (\ref{B2n})
evaluated for the solution (\ref{BH}) gives a finite contribution plus a
divergent piece,

\begin{eqnarray}
\int\limits_{\partial M}B_{2n}^{E} &=&\frac{\beta }{G_{n}}n\left( 1-\sigma
\right) \int\limits_{0}^{1}dt\,t\left( 1-t^{2}\left( 1-\sigma \right)
\right) ^{n-1}  \nonumber \\
&&+2(d-2)!\Omega _{d-2}\beta \kappa n\left[ r^{2}\int\limits_{0}^{1}dt\left(
\sigma +\left( t^{2}-1\right) r^{2}\right) ^{n-1}\right] ^{r=\infty }.
\label{B2nEval}
\end{eqnarray}
Analogously, the bulk term takes the form
\begin{eqnarray}
I_{2n+1}^{E} &=&(d-2)!\Omega _{d-2}\beta \kappa \left[
2nr^{2}\int\limits_{0}^{1}dt\left( t^{2}-1\right) \left( \sigma +\left(
t^{2}-1\right) r^{2}\right) ^{n-1}\right.  \label{BulkI} \\
&&\qquad \qquad \qquad \qquad +\left. \int\limits_{0}^{1}dt\left( \sigma
+\left( t^{2}-1\right) r^{2}\right) ^{n}\right] _{r=r_{+}}^{r=\infty },
\end{eqnarray}
which also has a divergent piece at infinity, plus a finite term coming from
the horizon. It it easy to check that the divergent pieces from both
expressions cancel out, leaving a finite Euclidean action,

\begin{equation}
I^{E}_{2n+1}=\frac{\beta }{G_{n}}nr_{+}^{2}\int\limits_{0}^{1}dt\left( \sigma
+\left( t^{2}-1\right) r_{+}^{2}\right) ^{n-1}+\frac{\beta }{G_{n}}n\left(
1-\sigma \right) \int\limits_{0}^{1}dtt\left( 1-t^{2}\left( 1-\sigma \right)
\right) ^{n-1},  \label{BetaF}
\end{equation}
where the first term comes from the horizon and the second from infinity.
The mass is then found to be

\begin{equation}
M=-\frac{\partial I_{2n+1}}{\partial \beta }=\frac{\sigma ^{n}-1}{2G_{n}}%
,  \label{Mass}
\end{equation}
while the entropy reads
\begin{eqnarray}
S &=&\left( 1-\beta \frac{\partial }{\partial \beta }\right) I_{2n+1}
\label{SBH} \\
&=&\frac{2\pi }{G_{n}}nr_{+}\int\limits_{0}^{1}dt\left(
1+t^{2}r_{+}^{2}\right) ^{n-1},
\end{eqnarray}
which, by means of an appropriate change of variable, leads to the
expression
\begin{equation}
S=\frac{2\pi }{G_{n}}n\int\limits_{0}^{r_{+}}dr\left( 1+r^{2}\right) ^{n-1}.
\label{SBHfinal}
\end{equation}
Note that, as expected, the term coming from infinity in Eq. (\ref{BetaF})
is identified as $-\beta M$, while the term from the horizon is the entropy,
as found by different methods \cite{Myers-Simon,DCBH,Disquito,scan,Ross}.

Analogously, for the topological black holes (\ref{BH}), the mass is

\begin{equation}
M=\frac{\Sigma _{\gamma }}{2\Omega _{d-2}G_{n}}\left[ \sigma ^{n}-\gamma
^{n}\right] ,  \label{Massgamma}
\end{equation}
where $\Sigma _{\gamma }$ is the volume of the base manifold, and the
entropy is given by

\begin{equation}
S=\frac{2\pi }{G_{n}}n\frac{\Sigma _{\gamma }}{\Omega _{d-2}}%
\int\limits_{0}^{r_{+}}dr\left( \gamma +r^{2}\right) ^{n-1},
\label{Stop}
\end{equation}
which is also in agreement with the Hamiltonian formalism \cite{aros}. Note
that the mass of the negative constant curvature configurations (locally AdS)
depends on the topology of the boundary, and is given by

\begin{equation}
M_{0}=-\frac{\Sigma _{\gamma }}{2\Omega _{d-2}G_{n}}\gamma ^{n},
\label{M0}
\end{equation}
which could be interpreted as the Casimir energy of the dual CFT, reflecting
the existence of the Weyl anomaly.

\section{Conserved Charges as Surface Integrals}

The action principle presented here allows to write the conserved charges
associated with asymptotic symmetry as surface integrals at infinity in a very
straightforward manner. By direct application of Noether's theorem, the
conserved current associated with the invariance under diffeomorphisms of the
Lagrangian $L_{2n+1}$ is given by (see Appendix A.2)
\begin{equation}
\ast J=-\Theta -I_{\xi }L_{2n+1}  \label{Jdiff}
\end{equation}
where $\Theta $ is the boundary term that comes from the variation of the
action on shell evaluated for a change in the fields induced by a
diffeomorphism, and $I_{\xi }$ is the contraction operator\footnote{The action
of the contraction operator $I_{\xi }$ over a $p$-form $\alpha
_{p}=\frac{1}{p!}\alpha _{\mu _{1}}\ldots _{\mu _{p}}dx^{\mu _{1}}\ldots
dx^{\mu _{p}}$ is given by $I_{\xi }\alpha _{p}=\frac{1}{(p-1)!}\xi ^{\nu
}\alpha _{\nu \mu _{1}}\ldots _{\mu _{p-1}}dx^{\mu _{1}}\ldots dx^{\mu
_{p-1}}$. In terms of this operator, the Lie derivative reads ${\cal L}_{\xi
}=dI_{\xi }+I_{\xi }d$.}. The field equations allows to write the current as an
exact form, $*J=dQ(\xi )$, and assuming suitable asymptotic conditions for the
fields, the conserved charge can be expressed as the surface integral

\begin{equation}
Q(\xi )=\int\limits_{\partial \Sigma }\left( I_{\xi }\theta ^{ab}\frac{%
\delta L_{2n+1}}{\delta R^{ab}}+I_{\xi }\theta ^{ab}\frac{\delta B_{2n}}{%
\delta \theta ^{ab}}+I_{\xi }e^{a}\frac{\delta B_{2n}}{\delta e^{a}}\right) .
\label{chargegeneral}
\end{equation}

This defines a conserved charge when the parameter $\xi $ is an asymptotic
Killing vector. In order to write (\ref{chargegeneral}) we have demanded
that the connection at the boundary be left unchanged by a displacement
along $\xi $, that is, ${\cal L}_{\xi }\omega =0$, which is not an
additional requirement. For the action (\ref{LCSAdS}), the charge is given by

\begin{equation}
Q(\xi )=\kappa n\int\limits_{\partial \Sigma }\int\limits_{0}^{1}dtt\epsilon
\left( I_{\xi }\theta e+\theta I_{\xi }e\right) \left( \widetilde{R}%
+t^{2}\theta ^{2}+t^{2}e^{2}\right) ^{n-1}.  \label{chargexi}
\end{equation}

As the black hole solutions (\ref{BH}) possess a timelike Killing vector $%
\partial _{t}$, the mass can also be evaluated as $Q(\partial _{t})=M$. For
instance, for the spherically symmetric solution ($\gamma =1$), Eq. (\ref
{chargexi}) yields (see Appendix A.3)

\begin{eqnarray}
Q(\partial _{t}) &=&2\kappa n\int\limits_{\partial \Sigma
}\int_{0}^{1}dt~t\epsilon _{01m_{1}...m_{d-2}}(\theta
_{t}^{01}e^{m_{1}}+e_{t}^{0}\theta ^{1m_{1}})[1-t^{2}(1-\sigma )]^{n-1}%
\tilde{e}^{m_{2}}...\tilde{e}^{m_{d-2}}  \nonumber \\
&=&\frac{n\left( 1-\sigma \right) }{G_{n}}\int\limits_{0}^{1}dtt\left(
1-t^{2}\left( 1-\sigma \right) \right) ^{n-1}=\frac{1}{2G_{n}}\left( \sigma
^{n}-1\right) ,  \label{Q(dt)}
\end{eqnarray}
which agrees with the previous result (\ref{Mass}). In general, for all
values of $\gamma $, one finds

\begin{equation}
Q\left( \partial _{t}\right) =\frac{\Sigma _{\gamma }}{2\Omega _{d-2}G_{n}}%
\left[ \sigma ^{n}-\gamma ^{n}\right] ,  \label{Qtop}
\end{equation}
in agreement with (\ref{Massgamma}). We should note that the integrand in
the first expression for $Q(\partial _{t})$ in (\ref{Q(dt)}) does not depend
on $r$. Therefore, the mass could be obtained integrating on a surface $%
\partial \Sigma $ of any radius.

It is worth mentioning that although the mass has been computed following
two radically different approaches, the zero point energy (the mass of the
locally AdS solutions, $\sigma =0$) or Casimir energy is the same in both
cases.

\section{Discussion}

Imposing boundary conditions for the curvature instead of the metric in AdS
gravity also yields well-defined action principles for different AdS gravity
theories in even dimensions \cite {acotz1,acotz2,scan}. In fact, it can also be
shown that the same boundary term (\ref{B2n}) can render the Einstein-Hilbert
action in $2n+1$ dimensions finite,
\begin{equation}
I^{Reg}_{EH\;2n+1}=\int\limits_{M}\;\epsilon Re^{2n-1}+ \alpha
\int\limits_{\partial M}B_{2n} \;,  \label{FiniteEHAction}
\end{equation}
for an appropriate $\alpha$. In the same spirit, the present approach suggests
that it is possible to answer the same question in Gauss-Bonnet extended
gravity in odd dimensions through a similar set of boundary conditions.

The kind of theories considered here are generalizations of the
Einstein-Hilbert action containing terms with higher powers in the curvature.
For instance, the five-dimensional theory has a bulk term given by a particular
combination of the Einstein-Hilbert action with negative cosmological constant
and a Gauss-Bonnet term (\ref{GB}). The relevance of these theories has been
emphasized in brane-world scenarios (see, e.g., Refs. \cite{Braneworlds}), and
conserved charges have also been obtained in this case following different
approaches \cite{GB-Charges}. Further interesting issues related to
Chern-Simons AdS\ gravity have been explored in Refs. \cite{CS}

Here we have constructed a finite action principle for Chern-Simons AdS gravity
given by (\ref{FiniteAdSAction}). This is found requiring the action to attain
an extremum for the set of boundary conditions (\ref{BC}). This method does not
invoke a background configuration and it works for all odd dimensions. The
Euclidean continuation of the action was shown to correctly describe the black
hole thermodynamics in the canonical ensemble. It is worth mentioning that in
our case, the zero point energy (the mass of the locally AdS configurations
$M_{0}$) becomes fixed and depends on topology of the boundary as in
(\ref{M0}), in analogy with what occurs for the Einstein-Hilbert action with
the counterterms method \cite{Balasubramanian-Kraus,EJM}. This could be
interpreted as the Casimir energy of the dual CFT, signaling the presence of
the Weyl anomaly. These anomalies have been recently computed for Chern-Simons
AdS\ gravity in Ref. \cite{Banados-Schwimmer-Theisen}. The action principle
presented here also allows readily to construct background-independent
conserved charges as surface integrals associated with the asymptotic
symmetries, by direct application of Noether's theorem. Alternative conserved
charges have been obtained for Chern-Simons AdS\ gravity by different methods
in Refs. \cite {SS,F,Banados-Schwimmer-Theisen}. It would be interesting to
examine the relationship between these expressions, as well as to explore what
would be obtained through time-honored hamiltonian methods \cite{Time-Honored},
as well as with different recent approaches
\cite{Balasubramanian-Kraus,Recent}.

It is also worthwhile to observe that the action studied in this paper can be
regarded as a transgression form, which seems to be the ultimate reason behind
its good properties \cite{motz2}.

\appendix

\section{Appendices}

\subsection{The second fundamental form and the Gauss-Codazzi equations}

When one deals with local orthonormal frames $e^{a}=e_{\mu }^{a}dx^{\mu }$ on a
bounded manifold $M$, the role of the extrinsic curvature $K_{ij}$ is played by
the second fundamental form $\theta ^{ab}$, which can be defined as follows.
Let us consider a manifold $\bar{M}$ which is cobordant with $M$, i.e.,
$\partial M=\partial \bar{M}$. If $\bar{M}$ is endowed with a metric that
matches the metric of $M$ at the boundary, then the second fundamental form
$\theta ^{ab}$ is defined as the difference between the spin connections of $M
$ and $\bar{M}$ at the boundary (see e.g. \cite{eguchi}),
\begin{equation}
\theta ^{ab}=\left[\omega ^{ab}-\bar{\omega}^{ab}\right]_{\partial M}\;.
\label{sfdef}
\end{equation}
It is important to stress that $\bar{\omega}^{ab}$ is the spin connection of
the cobordant manifold $\bar{M}$, and has nothing to do with the usual concept
of background field. Indeed, in contrast with a background field, the
connection $\bar{\omega}^{ab}$ has no meaning in the bulk of $M$. Hence, as $%
\bar{\omega}^{ab}$ is defined only at the boundary, it can be naturally
identified as the connection for an auxiliary product manifold $\bar{M}$
that is cobordant to $M$.

The relationship between the second fundamental form and the extrinsic
curvature becomes clear using Gaussian coordinates, which can always be defined
in an open neighborhood near the boundary. In the vicinity of the boundary, the
metric of the spacetime $M$ can be written in Gaussian coordinates as
\begin{equation}
ds^{2}=dz^{2}+h_{ij}(z,x)dx^{i}dx^{j}\;,  \label{anymetric1}
\end{equation}
where $z$ is the coordinate normal to the boundary, defined by the surface
$z=0$. Analogously, the coordinates of the cobordant manifold $\bar{M}$ can be
chosen so that the metric reads
\begin{equation}
d\bar{s}^{2}=dz^{2}+h_{ij}(z=0,x)dx^{i}dx^{j}\;,  \label{productmetric}
\end{equation}
which matches the metric (\ref{anymetric1}) at the boundary. Using the
decomposition $a=\{1,\underline{i}\}$ for the tangent space indices and $\mu
=\{z,j\}$ for the world indices, the vielbein near the boundary of $M$ can be
chosen as
\begin{equation}
e^{1}=dz\;;\;e^{\underline{k}}=e_{j}^{\underline{k}}(z,x)dx^{j}\;,
\label{vielbein1}
\end{equation}
where $h_{ij}=\eta _{\underline{k}\,\underline{l}}e_{i}^{\underline{k}}
e_{j}^{\underline{l}}$. Analogously for $\bar{M}$ we choose $\bar{e}^{1}
=e^{1}$, and $\bar{e}^{\underline{i}}=e^{\underline{i}}(z=0,x)$. As the second
fundamental form is defined at the boundary, it has no components along $dz$,
i.e., $\theta ^{ab}=\theta _{j}^{ab}dx^{j}$. The corresponding components of
the spin connections $\omega ^{ab}$ and $\bar{\omega}^{ab}$ are obtained from
the vanishing of the torsion. Since $\bar{M}$ is a product manifold,
$\bar{\omega}^{ab}$ does not have normal components, and hence the only non
vanishing components are given by $\bar{\omega}^{\underline{%
i}\underline{j}}=\omega _{k}^{\underline{i}\underline{j}}(z=0,x)dx^{k}$,
which means that $\theta ^{\underline{i}\underline{j}}=\left. \left( \omega
^{\underline{i}\underline{j}}-\bar{\omega}^{\underline{i}\underline{j}%
}\right) \right| _{\partial M}=0$, and $\theta ^{1\underline{i}}=\omega
_{j}^{1\underline{i}}dx^{j}$, with $\omega _{j}^{1\underline{i}}=-\frac{1}{2}%
h_{jk,z}e^{\underline{i}k}$. On the other hand, since the extrinsic
curvature is defined as $K_{jk}=\nabla _{j}n_{k}$, where $n_{k}$ is normal
to the boundary whose only nonvanishing component is $n_{z}=1$, one obtains
that

\begin{equation}
K_{jk}=\partial _{j}n_{k}-\Gamma _{jk}^{\mu }n_{\mu }=\frac{1}{2}h_{jk,z}.
  \label{Kijdef}
\end{equation}
Therefore,
\begin{eqnarray}
\theta ^{1\underline{i}} &=&-K_{j}^{\;k}e_{k}^{\underline{i}}dx^{j}
\nonumber \\
\theta ^{\underline{i}\underline{j}} &=&0\;.  \label{secff}
\end{eqnarray}
expresses the relation between the second fundamental form and the extrinsic
curvature.

Note that the curvature two-form $R^{ab}$ can also be decomposed in tangent
and normal components defined by

\begin{eqnarray*}
R^{\underline{i}\underline{j}} &=&d\omega ^{\underline{i}\underline{j}%
}+\omega _{\;\underline{k}}^{\underline{i}}\omega ^{\underline{k}\underline{j%
}}+\omega _{\;1}^{\underline{i}}\omega ^{1\underline{j}}\;, \\
R^{1\underline{i}} &=&d\omega ^{1\underline{i}}+\omega _{\;\underline{k}}^{%
\underline{1}}\omega ^{\underline{k}\underline{i}}\;,
\end{eqnarray*}
which at the boundary can be expressed as
\begin{eqnarray*}
R^{\underline{i}\underline{j}} &=&d\overline{\omega }^{\underline{i}%
\underline{j}}+\overline{\omega }_{\;\underline{k}}^{\underline{i}}\overline{%
\omega }^{\underline{k}\underline{j}}+\theta _{\;1}^{\underline{i}}\theta ^{1%
\underline{j}}\;, \\
R^{1\underline{i}} &=&d\theta ^{1\underline{i}}+\overline{\omega }_{%
\underline{k}}^{\underline{i}}\theta ^{1\underline{k}}\;.
\end{eqnarray*}
This allows to write the Gauss-Codazzi equations in terms of the second
fundamental form

\begin{eqnarray}
R^{\underline{i}\underline{j}} &=&\widetilde{R}^{\underline{i}\underline{j}%
}+\left( \theta ^{2}\right) ^{\underline{i}\underline{j}}\;,  \label{GaussC}
\\
R^{1\underline{i}} &=&\tilde{D}\theta ^{1\underline{i}}\;,
\end{eqnarray}
where $\tilde{D}$ and $\widetilde{R}^{\underline{i}\underline{j}}$ are the
Lorentz covariant derivative, and the curvature two-form of $\partial M$%
, respectively. Hence, the decomposition (\ref{GaussC}) recovers the
well-known tensorial form of Gauss-Codazzi relations
\begin{equation}
R_{kl}^{ij}=\widetilde{R}_{kl}^{ij}-K_{k}^{i}K_{l}^{j}+K_{l}^{i}K_{k}^{j}
\label{GCtensor}
\end{equation}
where $R_{kl}^{ij}$ and $\widetilde{R}_{kl}^{ij}$ the Riemann tensors of the
bulk and the boundary metrics, respectively.

As an application, note that since the boundary $\partial M$ located at
fixed $z$, the components of $R^{ab}$ and $e^{a}$ along $dz$ do not
contribute to the expression (\ref{BTvar5}). Hence, using the Gauss-Codazzi
equations (\ref{GaussC}), the boundary term (\ref{theta}) reads
\begin{equation}
\Theta =4\kappa \int\limits_{\partial M}\epsilon _{1\underline{i}\underline{j%
}\underline{k}\underline{l}}\delta \theta ^{1\underline{i}}e^{\underline{j}%
}\left( \widetilde{R}^{\underline{k}\underline{l}}+\left( \theta ^{2}\right) ^{%
\underline{k}\underline{l}}+\frac{1}{3}e^{\underline{k}}e^{\underline{l}%
}\right) +\delta B_{4}\;,  \label{boundary41j}
\end{equation}
which by virtue of Eq. (\ref{secff}) it can be covariantized back as

\begin{equation}
\Theta =2\kappa \int\limits_{\partial M}\epsilon _{abcdf}\delta \theta
^{ab}e^{c}\left( \widetilde{R}^{df}+\left( \theta ^{2}\right) ^{df}+\frac{1}{3}%
e^{d}e^{f}\right) +\delta B_{4}\;.
\end{equation}
Therefore, the net effect is that the connection $\bar{\omega}^{ab}$ can be
regarded as being just a reference field fixed at the boundary, so that $\delta
\omega ^{ab}$ can be replaced by $\delta \theta ^{ab}$. Consistently, the
curvature at the boundary$\;\widetilde{R}^{\underline{i}\underline{j}}=%
\widetilde{R}^{\underline{i}\underline{j}}(\bar{\omega})$ possesses a
vanishing variation, so that the variation for the curvature two-form $R^{%
\underline{i}\underline{j}}$ is given by
\begin{equation}
\delta R^{\underline{i}\underline{j}}=D\delta \omega ^{\underline{i}%
\underline{j}}=\delta \left( \theta ^{2}\right) ^{\underline{i}\underline{j}%
}\;.
\end{equation}

Thus, in higher odd dimensions one proceeds in the same way. The surface
term $\Theta $ now reads
\begin{equation}
\Theta =\alpha _{2n}+\delta B_{2n}  \label{theta2n}
\end{equation}
where $\alpha _{2n}$ is the boundary term coming from the variation of the bulk
term in Eq. (\ref{Lovelock})\footnote{Note that in the last expression it is
easy to recognize the Lagrangian of CS-AdS gravity for the odd dimension right
below: $\alpha _{2n}=\kappa n\epsilon _{a_{1}...a_{2n+1}}\delta \omega
^{a_{1}a_{2}}L_{2n-1}^{a_{3}...a_{2n+1}}(R,e)$.},
\begin{equation}
\alpha _{2n}=\kappa n\int\limits_{0}^{1}dt\epsilon _{a_{1}...a_{2n+1}}\delta
\omega ^{a_{1}a_{2}}e^{a_{3}}R_{t}^{a_{4}a_{5}}...R_{t}^{a_{2n}a_{2n+1}},
\label{BTvar}
\end{equation}
which can be expressed as

\begin{equation}
\alpha _{2n}=\kappa n\int\limits_{0}^{1}dt\epsilon \delta \theta
e\sum\limits_{k=0}^{n-1}C_{k}^{n-1}\left( \widetilde{R}+t^{2}e^{2}\right)
^{n-1-k}\theta ^{2k}\;,
\end{equation}
with $C_{p}^{n}=\left(
\begin{array}{l}
n \\
p
\end{array}
\right) $. It is simple to see that the boundary condition (\ref{BC}) makes
it possible to integrate $B_{2n}$ from its variation as in the
five-dimensional case, since Eq. (\ref{BC}) allows to express $\alpha _{2n}$
as

\begin{eqnarray}
\alpha _{2n} &=&\kappa n\int\limits_{0}^{1}dt\epsilon \delta \theta
e\sum\limits_{k=0}^{n-1}C_{k}^{n-1}\left( \widetilde{R}+t^{2}e^{2}\right)
^{n-1-k}\theta ^{2k}\left( 1-t^{2k+1}\right)  \nonumber \\
&&+\kappa n\int\limits_{0}^{1}dt\epsilon \theta \delta
e\sum\limits_{k=0}^{n-1}C_{k}^{n-1}\left( \widetilde{R}+t^{2}e^{2}\right)
^{n-1-k}t^{2k+1}\theta ^{2k}.
\end{eqnarray}
After some algebraic manipulation, that includes identities similar to \cite
{reshuffling} in five dimensions, we integrate out the variation of $\theta $
and $e$ as

\begin{eqnarray}
\delta I_{2n+1} &=&\int\limits_{\partial M}\kappa
n\int\limits_{0}^{1}dt\epsilon
\sum\limits_{k=0}^{n-1}C_{k}^{n-1}\sum\limits_{l=0}^{n-1-k}C_{l}^{n-1-k}%
\widetilde{R}^{n-1-k-l}\int\limits_{0}^{t}dst^{2k+1}\delta \theta
^{2k+1}s^{2l}e^{2l+1}  \nonumber \\
&&+\kappa n\int\limits_{0}^{1}dt\epsilon
\sum\limits_{k=0}^{n-1}C_{k}^{n-1}\sum\limits_{l=0}^{n-1-k}C_{l}^{n-1-k}%
\widetilde{R}^{n-1-k-l}\int\limits_{0}^{t}dst^{2l+1}\theta
^{2l+1}s^{2k}\delta e^{2k+1}  \nonumber \\
&&+\delta B_{2n}.
\end{eqnarray}

Finally, the boundary term $B_{2n}$ can be integrated in a very concise form
as the double integral in Eq. (\ref{B2n}).

\subsection{Noether Theorem}

In order to fix the notation and conventions, here we briefly review
Noether's theorem. Consider a $d$-form Lagrangian $L(\varphi ,d\varphi )$,
where $\varphi $ denotes collectively a set of $p$-form fields. An arbitrary
variation of the action under a local change $\delta \varphi $ is given by
the integral of
\begin{equation}
\delta L=(E.O.M)\delta \varphi +d\Theta (\varphi ,\delta \varphi ),
\label{EOM}
\end{equation}
where E.O.M. stands for equations of motion and $\Theta $ is the
corresponding boundary term \cite{Ramond}. The total change in $\varphi $ ($%
\bar{\delta}\varphi =\varphi ^{\prime }(x^{\prime })-\varphi (x)$) can be
decomposed as a sum of a local variation and the change induced by a
diffeomorphism, that is, $\bar{\delta}\varphi =\delta \varphi +{\cal L}_{\xi
}\varphi $, where ${\cal L}_{\xi }$ is the Lie derivative operator. In
particular, a symmetry transformation acts on the coordinates of the manifold
as $\delta x^{\mu }=\xi ^{\mu }(x)$, and on the fields as $\delta \varphi $,
leading to a change in the Lagrangian given by $\delta L=d\Omega .$

Noether's theorem states that there exists a conserved current given by
\begin{equation}
\ast J=\Omega -\Theta (\varphi ,\delta \varphi )-I_{\xi }L,  \label{JNoether}
\end{equation}
which satisfies $d\ast J=0$. This, in turn, implies the existence of the
conserved charge
\begin{equation}
Q=\int\limits_{\Sigma }\;\ast J,
\end{equation}
where we assume a manifold with topology $M=R\times \Sigma $ and $\Sigma $ is
the spatial section of the manifold. If the Lagrangian is supplemented by a
boundary term $\alpha (\varphi, \overline{\varphi })$ that contains a
dependence on a fixed field $\overline{\varphi }$ at $\partial M$, the current
derived from the Noether theorem takes the form

\begin{equation}
\ast J^{\prime }=dQ+\frac{\delta \alpha }{\delta \overline{\varphi }}{\cal L}%
_{\xi }\overline{\varphi }.
\end{equation}
A sufficient condition to ensure the conservation of the current is taking $%
{\cal L}_{\xi }\overline{\varphi }=0$.

\subsection{Second fundamental form and the curvature of a black hole}

For the black holes considered here, the line element is given by
\begin{equation}
ds^{2}=\Delta (r)^{2}dt^{2}+\frac{dr^{2}}{\Delta (r)^{2}}+r^{2}d\Sigma
_{\gamma }^{2}  \label{BHmetric}
\end{equation}
where $d\Sigma _{\gamma }^{2}$ is the line element of the $(d-2)$%
-dimensional base manifold $\Sigma _{\gamma }$ with constant curvature $%
\gamma =\pm 1,0$. The corresponding vielbeins can be chosen as
\begin{eqnarray*}
e^{0} &=&\Delta (r)dt \\
e^{1} &=&\frac{dr}{\Delta (r)} \\
e^{n} &=&r\tilde{e}^{n}
\end{eqnarray*}
where $\tilde{e}^{n}$ corresponds to the vielbein of the base manifold $\Sigma
_{\gamma }$. If the cobordant manifold is endowed with a product metric that
matches (\ref{BHmetric}) at the boundary, we have
\begin{eqnarray*}
\theta ^{01} &=&\frac{1}{2}(\Delta ^{2})^{\prime }dt \\
\theta ^{1m} &=&-\frac{\Delta (r)}{r}e^{m}=-\Delta (r)\tilde{e}^{m}
\end{eqnarray*}
where the prime means radial derivative $\partial _{r}$. The curvature
two-form is given by

{\normalsize
\begin{eqnarray*}
R^{01} &=&-\frac{1}{2}(\Delta ^{2})^{\prime \prime }e^{0}e^{1}\;, \\
R^{0m} &=&-\frac{(\Delta ^{2})^{\prime }}{2r}e^{0}e^{m}\;, \\
R^{1m} &=&-\frac{(\Delta ^{2})^{\prime }}{2r}e^{1}e^{m}\;, \\
R^{mn} &=&(\gamma -\Delta ^{2})\tilde{e}^{m}\tilde{e}^{n}\;.
\end{eqnarray*}
}
\\
{\bf Acknowledgments}

We wish to thank M. Ba\~{n}ados, G. Barnich, M. Hassa\"{\i}ne, M. Henneaux, G.
Kofinas, K. Skenderis, and S. Theisen for helpful discussions. PM is grateful
to the Centro de Estudios Cient\'{\i}ficos (CECS) for the hospitality during
the development of this work, and to the International Center of Theoretical
Physics, ICTP-Trieste, which made his visits to Valdivia possible. RO thanks
Prof. M. Henneaux for kind hospitality at the Universit\'e Libre de Bruxelles.
This work was partially funded by FONDECYT grants 1010449, 1010450, 1020629,
1040921, 3030029, and 7010450 from FONDECYT. Institutional support to CECS from
Empresas CMPC is also acknowledged. CECS is a Millennium Science Institute and
is funded in part by grants from Fundaci\'{o}n Andes and the Tinker Foundation.


\begin{thebibliography}{99}
\bibitem{reshuffling}  In the expression (\ref{split}), the integration of
the term
\[
\epsilon _{abcde}\delta \theta ^{ab}\theta _{f}^{c}\theta ^{fd}e^{e}=\frac{1%
}{3}\epsilon _{abcde}\delta \left( \theta ^{ab}\theta _{f}^{c}\theta
^{fd}\right) e^{e}
\]
is possible by virtue of the following identity
\[
\epsilon _{abcde}\delta \theta ^{ab}\theta _{f}^{c}\theta
^{fd}e^{e}=\epsilon _{abcde}\theta ^{ab}\delta \theta _{f}^{c}\theta
^{fd}e^{e}+\frac{1}{2}\epsilon _{abcde}\theta ^{ab}\delta \theta ^{cd}\theta
_{f}^{e}e^{f}\;,  \label{Tcheby5}
\]
which comes from the invariance of the Levi-Civita tensor under Lorentz
transformations. Since the non-vanishing components of the second fundamental
form require the presence of a normal direction in tangent space (see Eq.
(\ref{secff})), the last term identically vanishes.
\end{thebibliography}
\end{document}